# Enhanced Q-factor and effective length silicon photonics filter utilizing nested ring resonators

Mahmoud A. Selim[1,2,*], Momen Anwar[3]

[1] *Ming Hsieh Department of Electrical and Computer Engineering, University of Southern California, CA 90007, USA*
[2] *Faculty of Engineering, Ain-Shams University, 1 Elsarayat St. Abbassia, Cairo, Egypt*
[3] *Si Ware Systems, Heliopolis, Cairo 11361, Egypt*

E-mail: ma059189@ucf.edu

## Abstract

In this study, we investigate a novel design of an on-chip nested coupled ring resonator to enhance the quality factor and the effective length of the resonator. The configuration consists of an open ring and racetrack resonator, with lengths of 340 μm and 184.4 μm, respectively, with a coupling ratio of 97/3. In this regard, the proposed nested cavity has been experimentally characterized and compared with a single cavity ring resonator. Our results show a significant improvement in the quality factor by a factor of four. This improvement in performance opens up exciting new possibilities for state-of-the-art applications, such as compact optical sensors and delay lines. Our proposed design represents a significant advancement in the field of integrated optics, and we believe that it has the potential to enable a broad range of applications in the future.

## 1. Introduction

Recently, silicon photonics technology has incited a flurry of interest in both scientific and industrial communities. What has motivated the efforts in this field is the promise of high-speed optical communication capabilities enabled by wavelength division multiplexing [1]. Moreover, photonic integrated circuits (PICs) benefit (CMOS) fabrication process [2]. In addition, such a platform offers exceptional on-chip sensing capabilities [3,4]. In turn, silicon photonics components have been exploited in for several state-of-the-art applications, for instance, on-chip photonic filters [5,6], modulators [7], telecommunication networks [8], neural networks [9], and label-free detection [10], ring laser gyroscope to mention a few. In particular, silicon photonics ring resonators play a vital role in most of the aforementioned applications due to their strong field enhancement on the resonance and narrowband wavelength selectivity [11]. In the nonlinear regime, ring resonators can be utilized for frequency comb generation and optical thermodynamics [12–14]. Furthermore, the variation of such configuration can be utilized in RF photonics [15], delay lines [16], wavelength division multiplexing (WDM) [1], and box-like filters [17]. The prospect of topological phases in ring cavities has also been investigated [18].

Quite recently, the generation of orbital angular momentum modes from micro rings has been explored [19].

Along these lines, silicon photonics ring cavities have been widely used in sensing and spectroscopy. In most of these applications, a high-quality factor is required. Yet, to enhance the quality factor of a resonator, both the losses and the coupling coefficients must be reduced. Practically speaking, the losses in such resonators are usually limited by the surface roughness and the intrinsic loss which is prespecified by fabrication technology capabilities. Another figure of merit for the ring resonators is the free spectral range (FSR) - the spacing between cavity resonances - which depends on the cavity's length. For many applications, a large FSR is preferred. This will translate to a very small cavity length which leads to high radiation loss and low-quality factor [20–22]. The use of a gain medium before the lasing threshold to enhance the quality factor has also been explored [23–25]. Yet, the gain medium introduces spontaneous emission noise and complexity into the whole system.

To this end, several configurations have been proposed to improve the quality factor or finesse, the ratio between the spectral range and the full width at half maximum (FWHM), and the quality factor [26–28], among which is the cascaded ring resonator arrangement in a serial or parallel manner. Although these arrangements will enhance the overall quality factor and the free spectral range due to the vernier effect, the fabrication imperfection may shift the resonance wavelength of one cavity, thus hindering its use in practical applications. An alternative design is to couple several ring cavities in a panda configuration [29]. Along these lines, hybrid integration has also been utilized to enhance the linewidth of semiconductor lasers [30]. Moreover, the omni-resonant operation has been also demonstrated [31,32]. Although that such configuration allows for a broad spectral range, the quality factor is not significantly enhanced. As such, of interest will be to develop a novel design that can enhance the quality factor and the free spectral range and be robust against fabrication process variation. Furthermore, it would be beneficial to improve the sensitivity. In this letter, a novel silicon photonics filter configuration is presented to enhance the finesse and the free spectral range. The configuration has been both numerically and experimentally characterized and compared with a conventional ring resonator. The results demonstrate the effectiveness of the proposed design.

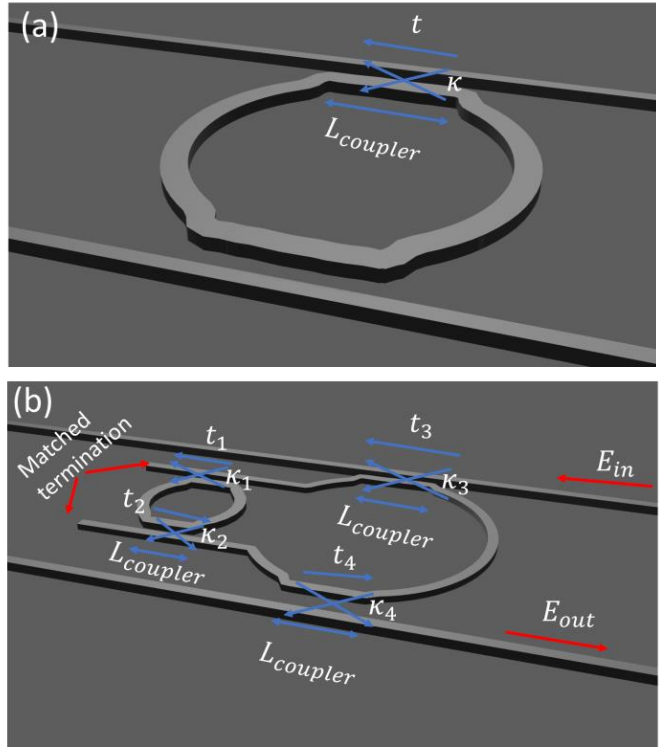

**Figure 1.** Schematics of (a) the single cavity configuration. (b) the nested cavity configuration.

## 2. Analysis of the nested cavity

We begin our analysis by considering a nested ring cavity configuration. Note that the schematic of the single ring cavity and the nested ring cavity is shown in figure 1(a) and 1(b), respectively. The field transfer function for the nested ring resonator can be written as

$$\frac{E_{out}}{E_{in}} = \frac{\kappa_3 \kappa_4 e^{j\beta L_3} H_c}{1 - t_3 t_4 H_c e^{j\beta L}}, \quad (1)$$

where $\kappa_n$ and $t_n$ are the cross and forward coupling coefficients of the coupler $n$, respectively. While $L_3$ is the length between the input and output coupler. In addition, $H_c$ is the transfer function of the nested cavity. Note that in the case of the single-ring cavity, $H_c = 1$. From here, the power transmission coefficient can be written as

$$T = \frac{(\kappa_1 \kappa_2 \kappa_3 \kappa_4)^2 / 2}{C - A_c \cos(\theta_c) + A_o \cos(\theta) - A_c A_o \cos(\beta \Delta L)} \quad (2)$$

here, $2C = 1 + A_c^2 + A_o^2$ with $A_c = \exp(\alpha L_c)\, t_1 t_2$ and $A_o = \exp(\alpha L)\, t_3 t_4 \kappa_1 \kappa_2$. Moreover, $\theta_c = \beta L_c$ and $\theta = \beta L$ with $L$ and $L_c$ are the lengths of the main and coupled cavities, and $\beta$ stands for the propagation constant in the cavity. In equation (2), the difference in the optical propagation distance between the two cavities is $\Delta L = L - L_c$. In addition, the maximum insertion loss for a nested cavity is given by [33]

$$IL = \frac{(\kappa_1 \kappa_2 \kappa_3 \kappa_4)^2}{(1 - A_c - A_o)^2}. \quad (3)$$

Moreover, using $\cos(x) \approx 1 - x^2$, the full width at half maximum can be derived as [33]

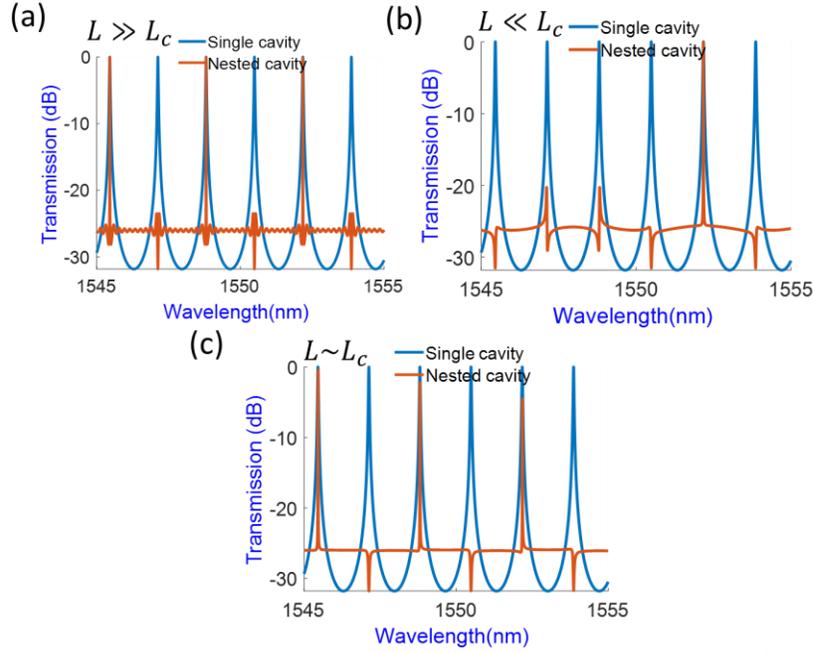

**Figure 2.** Nested cavity simulation at different lengths. a) $L = 34\ \mu m$ and $L_c = 340\ \mu m$ b) $L = 3400\ \mu m$ and $L_c = 340\ \mu m$ c) $L = 337.06\ \mu m$ and $L_c = 340\ \mu m$. In all cases the length of the single cavity is 340 μm. For simplicity, the cavity is assumed lossless in these simulations.

$$\Delta \lambda = \frac{(1 - A_c - A_o)\lambda_o^2/(n_g L)}{\pi\sqrt{A_c\left(\frac{L_c}{L}\right)^2 + A_o - A_c A_o\left(\frac{\Delta L}{L}\right)^2}}, \quad (4)$$

where $\lambda_o$ is the free space wavelength and $n_g$ is the group refractive index. It is worth emphasizing that the full width at half maximum for single ring resonator is $FWHM_s = \frac{(1-A)\lambda_o^2/(n_g L)}{\pi\sqrt{A}}$ and can be obtained from the previous formula by setting $t_1 = t_2 = 0$ and $\kappa_1 = \kappa_2 = 1$. In a straightforward manner, the quality factor can be derived as:

$$Q = \frac{\lambda}{\Delta\lambda} = \frac{\pi\sqrt{A_c\left(\frac{L_c}{L}\right)^2 + A_o - A_c A_o\left(\frac{\Delta L}{L}\right)^2}}{(1 - A_c - A_o)\lambda_o/(n_g L)}. \quad (5)$$

Meanwhile, the finesse of the cavity is usually defined as $F = FSR/FWHM$. Yet, in the case of the nested cavity, it is difficult to define a fixed FSR as there are many sub-resonance lines with different suppression ratios. To solve this issue, the free spectral range for the nested cavity is calculated for two full resonance modes [33]. Another parameter of interest in sensing applications and delay lines is the effective interaction length. Interestingly, the effective length is directly proportional to the quality factor of the cavity. The effective length of the resonator can be obtained from equation (1) by using the following approximation $\exp(-\alpha L_c) \approx 1 - \alpha L_c$. By doing so, one can obtain an approximate expression for the transmission:

$$T = \frac{(\kappa_1\kappa_2\kappa_3\kappa_4)^2\left(1 - \frac{2\alpha L_c t_1 t_2}{1 - t_3 t_4 \kappa_1 \kappa_2 - t_1 t_2}\right)}{(1 - t_3 t_4 \kappa_1 \kappa_2 - t_1 t_2)^2}. \quad (6)$$

From here, by comparing the latter formula with Beer-Lambert law for absorption ($T = c_o(1 - 2\alpha l)$), where $c_o$ is a constant that accounts for intrinsic loss, we can acquire the effective interaction length for the nested cavity as

$$L_{eff} = \frac{L_c t_1 t_2}{1 - t_3 t_4 \kappa_1 \kappa_2 - t_1 t_2}. \quad (7)$$

As expected, the effective length increases as the forward coupling coefficients increase. Interestingly, the total effective length of the nested cavity can be approximately obtained by multiplying the effective length of both cavities.

## 3. Parametric analysis

To illustrate these results, numerical simulations have been carried out to compare the nested and single cavities. In general, the regimes of operations for nested cavities could be divided into three regimes. First is when we have $L \gg L_c$, while the second is $L \sim L_c$ and the last regime is $L \ll L_c$. Figure 2 depicts the response of these three regimes, in all cases, the nested cavity has a higher finesse and quality factor than the single cavity. Note that throughout this paper, unless otherwise mentioned, the waveguide used is a strip waveguide with 500 nm width and 220 nm height. This, in turn, leads to a group refractive index of 4.2 and an effective refractive index of 2.42. In addition, for simplicity, we set $\kappa_1^2 = \kappa_2^2 = \kappa_3^3 = \kappa_4^2 = \kappa_o = 0.05$. Figure 3 depicts the enhancement factor in the quality factor and the effective length for the nested cavity. As indicated from these figures, as the cross-coupling coefficient increases, the effective length decreases regardless of the length of the main or the nested cavity in full accord with equation (7) . On the contrary, the quality factor exhibits nontrivial features versus the cross-coupling coefficient. At $L \gg L_c$ [see figures. 3(a) and 3(b)], the quality factor increases with increasing the cross-coupling coefficient, as the numerator is approximately proportional to $\propto \sqrt{A_o}/(1 - A_o)$. On the other hand, when $L \sim L_c$ or $L \gg L_c$, the quality factor peaks then gradually declines. Meanwhile, the effective length exhibits a consistent pattern by always falling off with the increase in the cross-coupling coefficient.

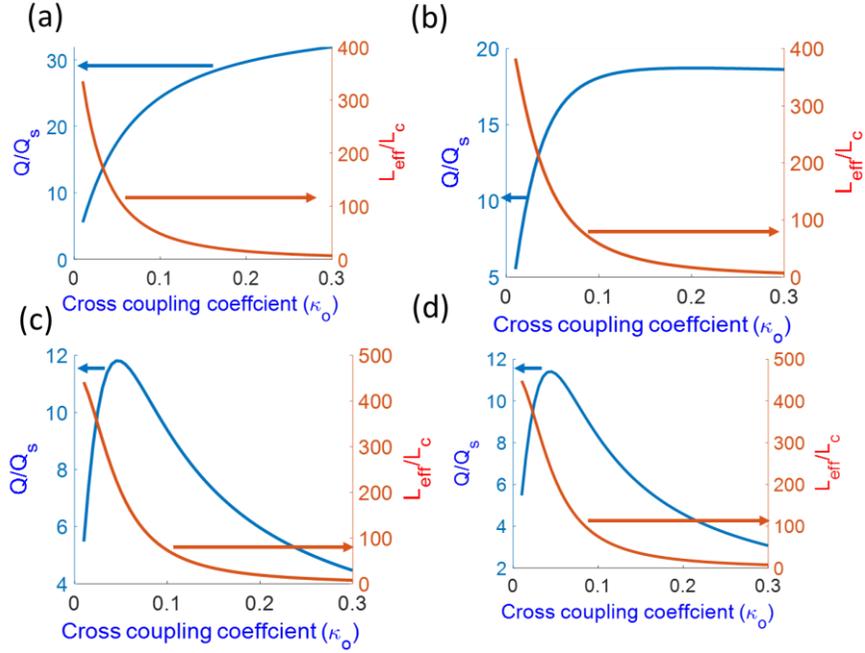

**Figure 3.** The enhancement factor in the quality factor and the effective interaction length at different main cavity dimensions. a) $L = 6.8$ mm b) $L = 3.4$ mm c) $L = 340$ μm. d) $L = 34$ μm . In all cases, the nested cavity length was taken to be $L_c = 184.4$ μm.

### 4. Experimental results

The described design was sent for fabrication at the Washington Nanofabrication Facility, University of Washington. This fabrication facility performs prototypes of passive silicon photonics by Electron Beam Lithography (EBL) [34]. The fabrication process can be summarized in the following steps [35]. A piece of SOI wafer is selected. Then, the resist layer is inserted by spinning and baking a negative photo-resistive material. Afterward, the wafer is exposed to the EBL. Next, the resist is developed by immersion in a hydroxide (TMAH) for a few minutes. Then, the wafer is submitted to a rinse and dry stage. Subsequently, the unexposed areas are removed by plasma etching. After that, the oxide is deposited over the entire chip. Finally, the chip is diced. Due to fabrication issues, there are several aspects that include mismatches between the designed and the fabricated devices. Mainly these manufacturing errors generate variations mainly in the thickness and width of the waveguides. As an approximation, it is common to associate the fabrication variability of a given geometrical parameter with a Gaussian distribution, where the nominal value corresponds to the distribution mean, and the fabrication error is expressed by the standard deviation. The thickness variation associated with the SOI wafer provided by the supplier is shown in figures 4(a) and 4(b). The results elaborate the corner analysis, which accounts for the width and thickness variability as depicted in figure 4 inset. More details about the fabrication process are shown in figure 4(c). Due to financial constraints, we were unable to obtain scanning electron microscope pictures for our chip through a multi-project wafer run (MPW). The silicon waveguide possesses a width of approximately 500 nm, complemented by a height of around 220 nm, and is encased within a silicon dioxide cladding. To illustrate the simulation results, nested cavity and single cavity configurations have been fabricated. The designs have been measured using the automated process at the University of British Colombia (UBC). The configuration compromises an open ring and racetrack resonators of lengths of 340 μm and 184.4 μm, as shown in figure 1(b). The coupler has a length $L_{coupler}$ of approximately 55 μm, with the gap width carefully maintained at 0.2 μm. Consequently, the coupling coefficient is deliberately and consistently designed to be 97/3. As expected, the nested cavity exhibits a much longer free spectral range than the single cavity, as shown in figures. 5(a) and 5(b). The full width at half maximum for the proposed configuration has also enhanced from 0.8 nm to 0.2 nm. Note that the nested cavities have higher insertion loss due to longer propagation distance in the nested cavity, as clarified by equation (7). Interestingly, the insertion loss of the nested cavity is higher than the single cavity, as shown in figure 5(c). This is in good agreement with equation (7) as the nested cavity magnifies the total effective length, and hence, the intrinsic losses [see figures 5(a)-5(c)]. Such features can be exploited in conjunction with cavity-enhanced techniques in biosensing applications [15,38–40]. Moreover, in conjunction with an optical modulator, it can provide a variable delay line. Note that based on the design parameters, the FWHM and FSR can be further enhanced based on the configuration and the regime. For instance, if the design constraints demand a high FSR, the vernier effect can be used with two cavities having approximately equal lengths $L{\sim}L_c$. Note that the nested cavity is able to provide a moderate enhancement factor in the Q-factor (which could be enhanced further by appropriate design) with only two cavities, as shown in Table 1. In this regard, the quality factor of the nested cavity, obtained using Q=λ/Δλ, is approximately 78250, whereas the quality factor of the single cavity is approximately 21438. The disparity between the simulation and experimental results is primarily attributed to the fact that surface roughness loss has a more pronounced effect on the quality factor of the nested cavity compared to the single cavity. The nested cavity enhances the effective path length by compelling light to undergo multiple round trips inside it. This effect is roughly equivalent to the number of round trips within the main cavity multiplied by the number of round trips in the single configuration. However, this enhancement also results in higher insertion loss in the case of the nested cavity. These characteristics position the nested cavity as a robust solution for optical communication and sensing applications.

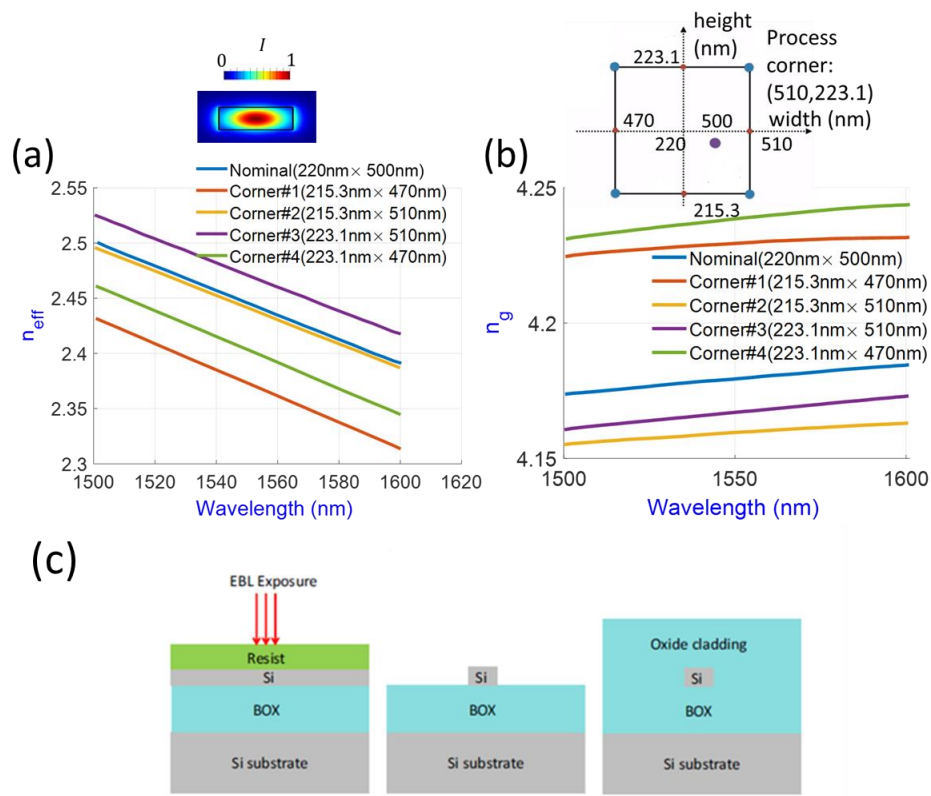

**Figure 4.** Corner analysis of the a) effective index and b) group index for the TE mode. The inset shows the normalized TE electric field intensity profile in the strip waveguide. c) The steps of fabrication process.

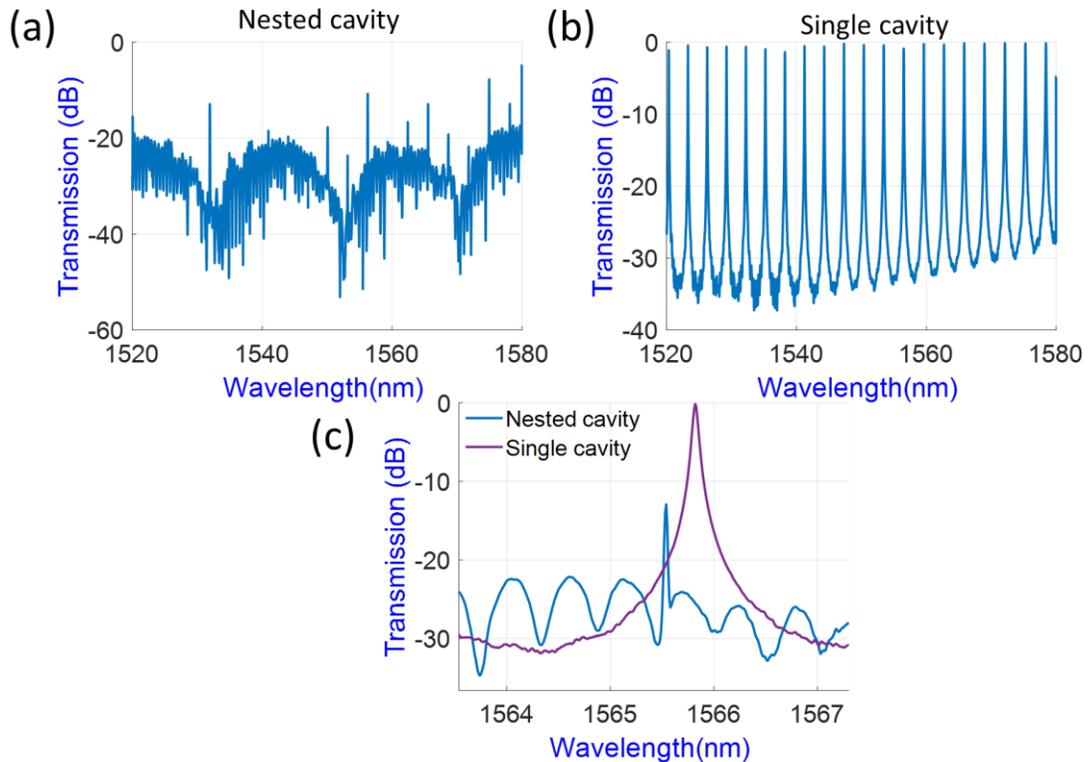

**Figure 5.** Experimental results for the two configurations. a) Nested cavity. b) Single cavity. c) Zoomed version for the previous plots around 1565 nm.

**Table 1.** Comparison between different configurations

| Configuration | Enhancement in Q-factor | Number of cavities | Reference |
|---|---|---|---|
| Nested Cavity | 4 ($Q_{nested} = 78250, Q_{single} = 21438$) | 2 | |
| Second-order Cascaded ring resonator | 2 | 2 | [36] |
| Integrated Fabry-Perot Cavity | 11 | 3 | [37] |

## 5. Conclusion

In conclusion, a comprehensive study of the nested cavity resonator has been presented. The proposed silicon photonics filter allows for the precise engineering of key cavity parameters. Our innovative design has demonstrated a remarkable four-fold increase in quality factor compared to the conventional ring cavities, making it a highly promising solution for on-chip spectral filters with high finesse ratios. Furthermore, this novel filter has the potential to enable a range of advanced applications, including slow light and variable delay lines. The exceptional performance and versatility of our silicon photonics filter make it a valuable addition to the toolkit of researchers and engineers working in the field of photonics. With further development, this technology could pave the way for new advances in a variety of fields, including telecommunications, sensing, and quantum computing.

## Data availability statement

All data that support the findings of this study are included within the article (and any supplementary files).

## Acknowledgments

This work was partially supported by Si-Ware Systems; the authors also would like to thank Lukas Chrostowski for facilitating the fabrication process and the measurements through the SiEPIC program.

## Disclosures

The authors declare no conflicts of interest.

## ORCID iDs

M. A. Selim https://orcid.org/0000-0001-5757-4089